\documentclass[11pt]{article}

\usepackage{amsmath}
\usepackage{graphicx}
\usepackage{amsfonts}
\usepackage{amssymb}
\usepackage{epsfig}
\usepackage{color}
\usepackage{psfrag}
\usepackage{epstopdf}

\usepackage{caption}
\usepackage{subcaption}

\newcommand{\EQ}{\begin{equation}}
\newcommand{\EN}{\end{equation}}
\newcommand{\be}{\begin{equation}}
\newcommand{\ee}{\end{equation}}
\newcommand{\bea}{\begin{eqnarray}}
\newcommand{\eea}{\end{eqnarray}}

\begin{document} \setcounter{page}{0}
\newpage
\setcounter{page}{0}
\renewcommand{\thefootnote}{\arabic{footnote}}
\newpage
\begin{titlepage}
\begin{flushright}
\end{flushright}
\vspace{0.5cm}
\begin{center}
{\large {\bf Critical exponents at the Nishimori point}}\\
\vspace{1.8cm}
{\large Gesualdo Delfino}\\
\vspace{0.5cm}
{\em SISSA -- Via Bonomea 265, 34136 Trieste, Italy}\\
{\em INFN sezione di Trieste, 34100 Trieste, Italy}\\
\end{center}
\vspace{1.2cm}

\renewcommand{\thefootnote}{\arabic{footnote}}
\setcounter{footnote}{0}

\begin{abstract}
\noindent
The Nishimori point of the random bond Ising model is a prototype of renormalization group fixed points with strong disorder. We show that the exact correlation length and crossover critical exponents at this point can be identified in two and three spatial dimensions starting from properties of the Nishimori line. These are the first exact exponents for frustrated random magnets, a circumstance to be also contrasted with the fact that the exact exponents of the Ising model without disorder are not known in three dimensions. Our considerations extend to higher dimensions and models other than Ising.
\end{abstract}
\end{titlepage}

\newpage

That of systems with quenched disorder is a notoriously challenging problem of statistical physics \cite{Nishimori_book}. It gives rise to a pattern of critical behavior of which the Ising model with randomly distributed ferromagnetic and antiferromagnetic bonds \cite{EA} provides a basic illustration. In three dimensions, the addition of weak randomness \cite{Harris} to the pure ferromagnet leads to a new renormalization group fixed point which can be studied perturbatively and is not yet sensitive to the specific type of disorder (the randomly site diluted ferromagnet falls in the same universality class)  \cite{Cardy_book}. As the fraction of antiferromagnetic bonds is increased, however, frustration starts to matter and eventually leads to the presence of the spin glass phase and of strong disorder fixed points. The critical properties in this region of the phase diagram remained traditionally inaccessible to analytical methods and have been the subject of extensive numerical work \cite{BY,KR}. Recently, however, it has been shown how, in two dimensions, conformal invariance can be exploited within the scattering framework to gain exact access to random fixed points of any disorder strength and reveal, in particular, some peculiar features of the relation of random criticality to universality \cite{random,colloquium}. 

The Nishimori point \cite{Nishimori_book,Nishimori,LdH1,LdH2} is the first strong disorder fixed point encountered as frustration is increased along the boundary of the ordered phase, and is located on a  specific line in the phase diagram for which some properties are known exactly \cite{Nishimori_book,Nishimori}. The unusual character of these properties has complicated the interpretation of their implications for the critical behavior at the Nishimori point. In this paper we show how, once further insight about these properties has been gained, it leads to the exact identification of the correlation length and crossover critical exponents at the Nishimori point in two and three dimensions. The rather accurate numerical estimates available nowadays for these exponents are in excellent agreement with the exact values. Our considerations extend to higher dimensions and to models other than Ising. The latter point, in particular, adds new insight on the interplay between randomness and internal symmetry already singled out by exact scattering theory in two dimensions.

\begin{figure}[t]
\centering
\includegraphics[width=7cm]{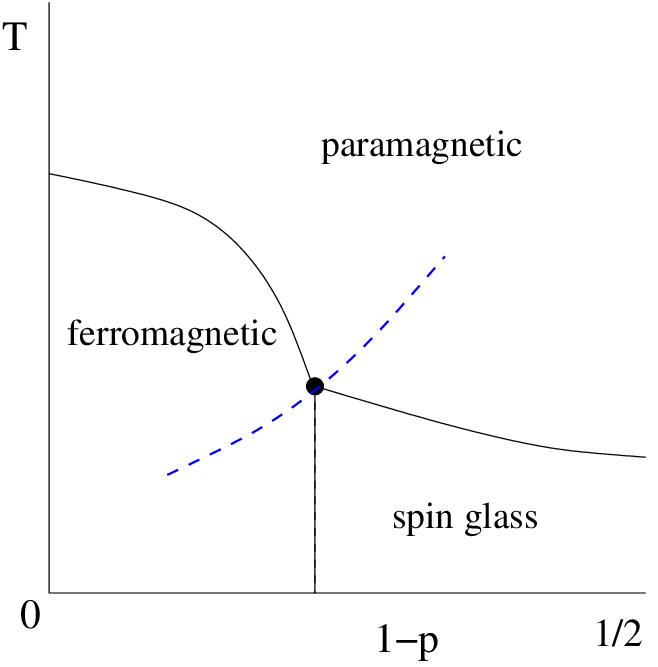}
\caption{Phase diagram of the $\pm J$ random bond Ising model. The Nishimori line (dashed) crosses the ferromagnetic phase boundary at the Nishimori multicritical point (dot). In two dimensions the spin glass phase is not observed for $T>0$.}
\label{pd}
\end{figure}

\vspace{.2cm}
The random bond Ising model \cite{EA} corresponds to the Hamiltonian
\EQ
H=-\sum_{\langle i,j\rangle}J_{ij}\sigma_i\sigma_j\,,
\label{lattice}
\EN
where $\sigma_i=\pm 1$ is the spin variable at site $i$ of a regular lattice, $\sum_{\langle i,j\rangle}$ denotes the sum over nearest neighbors, and disorder is introduced through random couplings $J_{ij}$ drawn from a probability distribution $P(J_{ij})$. A basic choice is the bimodal distribution
\EQ
P(J_{ij})=p\,\delta(J_{ij}-1)+(1-p)\,\delta(J_{ij}+1)\,,
\label{bimodal}
\EN
where $1-p$ is the fraction of antiferromagnetic bonds. The phase diagram emerging from numerical studies \cite{BY,KR} is shown in Figure~\ref{pd}. The ordered phase of the pure ferromagnet ($p=1$) persists as the disorder strength $1-p$ is increased, until the transition temperature vanishes. A multicritical point is present on the boundary of this ferromagnetic phase and, in three dimensions, another phase boundary separating the paramagnetic phase from a low temperature spin glass phase originates from the multicritical point. In two dimensions the spin glass phase is not observed for $T>0$. On bipartite lattices the phase diagram is symmetric under $p\to 1-p$.

For a class of probability distributions $P(J_{ij})$ allowing for a gauge symmetry, Nishimori showed that some exact results \cite{Nishimori_book,Nishimori} can be obtained along a line on the $p$--$T$ plane known as Nishimori line. This line crosses the plane from the ferromagnetic phase at low temperature to the paramagnetic phase at high temperature, and for the distribution (\ref{bimodal}) reads
\EQ
\tanh(1/T)=2p-1\,.
\label{NL}
\EN
The coincidence of the ferromagnetic order parameter $M$ with the spin glass order parameter $Q$ is an identity that can be derived on the Nishimori line \cite{Nishimori_book,Nishimori}, implying that the latter cannot enter a spin glass phase. Moreover, it has been argued \cite{LdH1,LdH2} -- and is supported by the numerical studies -- that in any dimension the Nishimori line is left invariant by renormalization group transformations, and that the point at which this line crosses the boundary of the ferromagnetic phase coincides with the multicritical point on this boundary. The multicritical point is then also known as the Nishimori point. 

Quenched disorder amounts to the fact that the average over the random variables $J_{ij}$ is taken on the free energy $F=-\ln Z$, where 
\EQ
Z=\sum_{\{\sigma_{i}\}}e^{-H/T}
\label{Z}
\EN
is the partition function of the system with assigned values of the random variables. Theoretically, it is convenient to use the identity 
\EQ
\ln Z=\lim_{n\to 0}\frac{Z^n-1}{n}\,,
\EN
by which the average over disorder has the effect of coupling $n\to 0$ replicas of the system with partition function $Z$. In the scaling limit close to second order transition points, the coupled replicas are described by an Euclidean field theory with reduced Hamiltonian ${\cal H}_n$. The averages $\langle\cdots\rangle$, performed with the weight $e^{-{\cal H}_n}$ for $n\to 0$, realize both the average over spin configurations and that over disorder for the original system (\ref{lattice}). In the following the limit $n\to 0$ will be understood.

Around the multicritical Nishimori point in $d$ spatial dimensions, the scaling reduced Hamiltonian can be written as
\EQ
{\cal H}_n={\cal H}_n^\textrm{FP}+g_1\int d^dx\,\varepsilon_1(x)+g_2\int d^dx\,\varepsilon_2(x)\,,
\label{scaling}
\EN
where ${\cal H}_n^\textrm{FP}$ is the reduced Hamiltonian of the renormalization group fixed point associated to the multicritical point, the $g_i$'s are couplings measuring the deviations from the multicritical point in the $p$--$T$ plane, and the $\varepsilon_i$'s are operators invariant under the spin reversal symmetry of the system; these operators are relevant in the renormalization group sense, namely have scaling dimensions $X_i<d$; our convention is
\EQ
X_1<X_2\,.
\label{convention}
\EN
Since ${\cal H}_n$ is dimensionless, the couplings scale as
\EQ
g_i\sim\xi^{-y_i}\,,
\label{gi}
\EN
where $\xi$ is the correlation length and
\EQ
y_i=d-X_i\,.
\label{yi}
\EN 
The theory (\ref{scaling}) describes the renormalization group trajectories flowing out of the multicritical point and propagating in the $p$--$T$ plane. Each such a trajectory is identified by a value of the dimensionless parameter
\EQ
\eta=g_1\,|g_2|^{-\phi}\,,
\label{eta}
\EN 
where
\EQ
\phi=\frac{y_1}{y_2}
\label{phi}
\EN
is the crossover exponent of the multicritical point. The Nishimori line corresponds to two trajectories -- one in the paramagnetic phase, the other in the ferromagnetic phase -- with values of $\eta$ that we will denote $\pm\eta_N$. 

When the multicritical point is approached along a trajectory, the correlation length diverges as
\EQ
\xi\simeq a_i(\eta)\,|g_i|^{-\nu_i}\,,
\label{xi}
\EN
where 
\EQ
\nu_i=\frac{1}{y_i}
\label{nui}
\EN
are the correlation length exponents at the multicritical point, and $a_i(\eta)$ are critical amplitudes. 

The singular part of the internal energy per unit volume is given by
\EQ
\frac{E_\textrm{sing}}{V}=\frac{T}{V}\,\langle{\cal H}_n\rangle=b(\eta)\,\xi^{-d}\simeq c_i(\eta)\,|g_i|^{d\nu_i}\,, 
\label{E_sing}
\EN 
where we call $c_i(\eta)$ the critical amplitudes which follow from the use of (\ref{xi}). On the other hand, the internal energy along the Nishimori line -- let us call it $E_N$ -- can be determined exactly on any lattice \cite{Nishimori_book,Nishimori}. For the bimodal disorder distribution (\ref{bimodal}) it reads
\EQ
E_N=N_B(1-2p)\,,
\label{E_Nishimori}
\EN 
where the total number of bonds $N_B$ corresponds to the volume $V$ times a lattice dependent factor. This result exhibits a property that holds also for the other disorder distributions allowing for the gauge symmetry: the internal energy is a regular function on the Nishimori line in spite of the fact that this line yields the multicritical point for some lattice dependent value of $p$. At first sight, this peculiar property could be attributed to the vanishing of the singular part (\ref{E_sing}) on the Nishimori line, namely to the vanishing of the critical amplitudes $c_i(\pm\eta_N)$. However, critical amplitudes are nonuniversal (i.e. lattice dependent) quantities, while $E_N$ is regular on any lattice. Explaining this regularity by the vanishing of $c_i(\pm\eta_N)$ would amount to say that these amplitudes take the same value (zero) on any lattice, thus contradicting their lattice dependence. The remaining possibility is that (\ref{E_sing}), which in general is the singular part of the internal energy, happens to be regular in this case. This requires
\EQ
d\nu_i=k_i(d)\,,
\label{quantization}
\EN
or equivalently
\EQ
X_i=\frac{k_i(d)-1}{k_i(d)}\,d\,,
\label{Xi}
\EN
where $k_i(d)$ are positive integers. The universality of the critical exponents $\nu_i$ then matches the lattice independence of the regularity of $E_N$. 

Some values of the integers $k_i$ can be ruled out exploiting the known fact that the specific heat is finite on the Nishimori line \cite{Nishimori_book,Nishimori}, and then also at the multicritical point. When this point is approached along the trajectory $g_2=0$, the singular (in generic cases) part of the specific heat behaves as
\EQ
C_\textrm{sing}\sim |g_1|^{-\alpha_1}\,,
\label{C_sing}
\EN
with the critical exponent
\EQ
\alpha_1=(d-2X_1)\nu_1=2-k_1(d)\,,
\label{alpha}
\EN
and absence of divergence at the multicritical point amounts to $\alpha_1<0$; the case $\alpha=0$ is excluded since it leads to a logarithmic divergence\footnote{Since the specific heat is proportional to the second derivative of the free energy with respect to the temperature, for a generic critical point we have $C_\textrm{sing}\sim\int d^dx\,\langle\varepsilon(x)\varepsilon(0)\rangle_\textrm{conn}$, where $\varepsilon$ is the energy density field. The short distance behavior $|x|^{-2X_\varepsilon}$ of the correlator in the integrand yields $C_\textrm{sing}\sim\ln (\xi/r_0)$ when $X_\varepsilon=d/2$ (i.e. $\alpha=0$), with $r_0$ a short distance cutoff. $C_\textrm{sing}$ then diverges logarithmically in the critical limit $\xi\to\infty$.}, as for the pure Ising model in $d=2$ \cite{Onsager}. Together with (\ref{convention}), the requirement of negative $\alpha_1$ implies
\EQ
k_2(d)>k_1(d)\geq 3\,.
\label{ki}
\EN

While (\ref{quantization}) does not uniquely fix the exponents $\nu_i$, it allows for them only a discrete set of values. Comparison with the numerical estimates of these exponents (Table~\ref{data}) selects the integers
\bea
&& k_1(2)=3\,,\hspace{1cm}k_2(2)=8\,,\label{k2d}\\
&& k_1(3)=3\,,\hspace{1cm}k_2(3)=5\,,\label{k3d}\\
&& k_1(4)=3\,,
\eea 
which indeed satisfy (\ref{ki}).

\begin{table}[t]
\begin{center}
\begin{tabular}{c|c|c|c|c|c}
\hline 
\hspace{-.5cm} Date & $y_1(2)$ & $y_2(2)$ & $y_1(3)$ & $y_2(3)$ & $y_1(4)$ \\ 
\hline \hline
1991 \cite{Singh} & $$ & $$ & $1.17(11)$ & $0.63(11)$ & \\ 
1996 \cite{SA} & $0.75(7)$ & $$ & $$ & $$ & $1.25(15)$\\
1999 \cite{Reis} & $$ & $\approx 0.25$ & $$ & $$ & \\
2001 \cite{HPP} & $0.75(2)$ & $$ & $$ & $$ & \\
2002 \cite{MC} & $0.67(3)$ & $0.25(3)$ & $$ & $$ & \\
2006 \cite{deQueiroz} & $0.67(1)$ & $$ & $$ & $$ & \\
2006 \cite{PHP} & $0.67(1)$ & $\sim 0.3$ & $$ & $$ & \\
2007 \cite{HPtPV_3d} & $$ & $$ & $1.02(5)$ & $0.61(2)$ & \\
2008 \cite{HPtPV} & $0.655(15)$ & $0.250(2)$ & $$ & $$ & \\
2009 \cite{PtPV} & $0.66(1)$ & $0.250(2)$ & $$ & $$ \\
2009 \cite{deQueiroz2} & $0.65(2)$ & $$ & $$ & $$ & \\
2014 \cite{WQZ} & $0.642(22)$ & $$ & $$ & $$ & \\
2024 \cite{Chen} & $0.67(1)$ & $$ & $$ & $$ & \\
\hline
present work & $2/3$ & $1/4$ & $1$ & $3/5$ & $4/3$ \\
\hline
\end{tabular} 
\caption{Numerical estimates of the inverse $y_i(d)$ of the correlation length exponents at the Nishimori point of the $d$-dimensional random bond Ising model together with, in the last line, their exact values.
}
\label{data}
\end{center}
\end{table}

It is worth observing that we deduced the conditions (\ref{quantization}) and (\ref{ki}) for the critical exponents $\nu_i$ starting from results known for the Nishimori line, which can be defined for specific disorder distributions allowing for a gauge symmetry. On the other hand, universality of the critical exponents implies that (\ref{quantization}) and (\ref{ki}) more generally hold for the disorder distributions which give rise to the multicritical point even if they do not allow for the gauge symmetry (this happens, for example, for a deformation of the bimodal distribution (\ref{bimodal}) to the case in which the two values of $J_{ij}$ are not equal in modulus). We will refer to the multicritical point as the ``Nishimori point" also in this more general case.

The above results lead to some additional considerations. In the first place, the fact that $k_1(d)=3$ for $d=2,3,4$ suggests the conjecture
\EQ
\nu_1=\frac{3}{d}
\label{nu1}
\EN
up to the upper critical dimension $d_c$. Since for $d\geq d_c$ the exponent $\nu_1$ has to take its mean field value $1/2$, (\ref{nu1}) yields $d_c=6$, which is indeed the known value originating from the presence of a cubic term in the Landau-Ginzburg Hamiltonian \cite{HLC,CL}. In $6-\varepsilon$ dimensions (\ref{nu1}) yields $\nu_1\simeq(1+\varepsilon/6)/2$, which differs from the result $(1+5\varepsilon/6)/2$ of the $\varepsilon$-expansion of \cite{CL} (see also \cite{LdH2}). The $\varepsilon$-expansion at the multicritical point has been known to be problematic because for $O(N)$ symmetry it yields complex critical exponents above the Ising value $N=1$, in spite of the fact that in the mean field regime $d>6$ the transition remains continuous with real exponents \cite{BY,CL}.

It was shown in \cite{ON} that a disordered XY model ($O(2)$ symmetry) can be defined which possesses a Nishimori line exhibiting the same properties holding for the Ising model. In particular, regular internal energy and nondivergent specific heat on any lattice \cite{ON} imply that our equations (\ref{quantization}) and (\ref{ki}) continue to hold at the multicritical point of the XY model. The numerical estimates $y_1=0.93(3)$ and $y_2=0.56(3)$ obtained in \cite{AV} for $d=3$ identify the integers (\ref{k3d}) as in the Ising case. The lower accuracy of the estimates of \cite{AV} for the XY model with respect to those of \cite{HPtPV_3d} for Ising can be traced back, in particular, to the smaller system size used in the Monte Carlo simulations; in any case, the estimate $\phi=1.7(1)$ obtained in \cite{AV} for the crossover exponent has the exact value $5/3$ within error bars. 

The fact that the Ising model and the XY model share the values of $k_1(3)$ and $k_2(3)$ means that the scaling dimensions $X_1$ and $X_2$ take the same values for two models with different symmetry, and are in this sense {\it superuniversal}. This is less surprising once one recalls that scale invariant scattering \cite{colloquium,paraf} shows exactly that the random fixed points of two-dimensional $N$-state Potts models \cite{random,DT2} and $O(N)$ models \cite{DL_ON1,DL_ON2} possess superuniversal -- i.e. not dependent on the symmetry parameter $N$ -- scattering sectors, a property which has no counterpart in absence of disorder. It was observed in \cite{random} that this can explain the absence of sizable $N$-dependence of the correlation length critical exponent $\nu$ observed since \cite{CFL1} in numerical studies of the weakest disorder fixed point in the two-dimensional $N$-state Potts model. More generally, it was argued in \cite{colloquium} that the scattering results point to superuniversality of {\it some} critical exponents as a rather common feature of random fixed points in two and, possibly, higher dimensions. At least for the multicritical Nishimori point, the conditions (\ref{Xi}) and (\ref{ki}) shed light on how this phenomenon may occur. 

The lower bound $d\nu\geq 2$ was rigorously proved in \cite{CCFS} for correlation length exponents at critical points of a large class of disordered systems. The conditions (\ref{quantization}) and (\ref{ki}) show that at Nishimori points the stronger bound $d\nu_i\geq 3$ holds in a discretized form. In addition, the indications of the Ising and XY cases are that the latter bound tends to be saturated by $\nu_1$.

In two dimensions, the possibility of obtaining exact critical exponents for quenched disorder was expected after it was shown in \cite{random} that conformal invariance can be used to access random fixed points. Seeing how our present results can contribute to the identification of conformal field theories of random criticality in $d=2$ is one of the interesting goals for future investigations. In higher dimensions (below $d_c$), having exact exponents is more surprising, but we have seen how this is made possible by the exam of $d$-independent exact properties of the Nishimori line through the lens of the renormalization group and universality.

\vspace{.2cm}
In conclusion, we showed how the properties of the Nishimori line imply a discretization condition for the two correlation length critical exponents at the multicritical Nishimori point of frustrated random magnets, and how this condition allows the exact identification of both exponents for the random bond Ising model in two and three dimensions. The existing numerical estimates are in excellent agreement with the exact values. The discretization condition at the multicritical point holds in any dimension, for any internal symmetry and, by universality, also for disorder distributions not allowing the definition of a Nishimori line. Our finding that the disordered Ising and XY models in $d=3$ share the same correlation length exponents at the Nishimori point adds new insight on the properties of superuniversality in random criticality revealed by recent exact results in $d=2$. Numerical investigations of systems with other internal symmetries can certainly contribute to the elucidation of this nontrivial emergent phenomenon.

\end{document}